\documentclass[
  aps,
  prfluids,
]{revtex4-2}

\usepackage[utf8]{inputenc}

\usepackage{amsfonts}
\usepackage{booktabs}

\usepackage[hidelinks]{hyperref}

\usepackage{amsmath}
\usepackage{amsthm}
\usepackage{dcolumn}
\usepackage{graphics}
\usepackage{graphicx}
\usepackage{subfigure}
\usepackage{amssymb}
\usepackage{graphics}
\usepackage{graphicx}
\usepackage{epstopdf}
\usepackage[svgnames]{xcolor}
\usepackage{subfigure}
\usepackage{url}
\usepackage{soul}
\usepackage{bm}  
\newcommand{\me}[1]{\langle #1 \rangle}

\newcommand\dertt[1]{ \frac{\partial{ #1}}{\partial t} }

\newcommand{\dd}{\text{d}}
\newcommand{\DD}{\text{D}}
\newcommand{\dert}[1]{\frac{\dd #1}{\dd t}}
\newcommand{\Dert}[1]{\frac{\DD #1}{\DD t}}

\newcommand*{\numParticles}{N_\sub{p}}

\newcommand*{\gradient}{\bm{\nabla}}
\newcommand*{\laplacian}{\nabla^2}

\newcommand*{\sub}[1]{\text{#1}}

\newcommand*{\vort}{\bm{\omega}}  

\newcommand*{\vn}{\bm{u}_\sub{n}}
\newcommand*{\vs}{\bm{u}_\sub{s}}

\newcommand*{\ws}{\vort_\sub{s}}
\newcommand*{\rhon}{\rho_\sub{n}}
\newcommand*{\rhos}{\rho_\sub{s}}
\newcommand*{\nun}{\nu_\sub{n}}
\newcommand*{\nus}{\nu_\sub{s}}
\newcommand*{\pn}{p_\sub{n}}
\newcommand*{\ps}{p_\sub{s}}

\newcommand*{\On}{\Omega_\sub{n}}
\newcommand*{\Os}{\Omega_\sub{s}}
\newcommand*{\Sn}{S_\sub{n}}
\newcommand*{\Ss}{S_\sub{s}}

\newcommand*{\An}{\Dert{\vn}}
\newcommand*{\As}{\Dert{\vs}}

\newcommand*{\Ucf}{U_\sub{ns}}  
\newcommand*{\UcfVec}{\bm{U}_\sub{ns}}

\newcommand*{\vRMS}{u_\sub{rms}}
\newcommand*{\vnRMS}{\vRMS^\sub{(n)}}

\newcommand*{\mutualFriction}{\bm{f}_\sub{ns}}  
\newcommand*{\forcing}[1]{\bm{\Phi}_\sub{#1}}

\newcommand*{\xvec}{\bm{x}}

\renewcommand*{\ap}{a_\sub{p}}  
\newcommand*{\rhop}{\rho_\sub{p}}
\newcommand*{\taup}{\tau_\sub{p}}

\newcommand*{\tauetaN}{\tau_\eta^{\sub{(n)}}}
\newcommand*{\etaN}{\eta^{\sub{(n)}}}
\newcommand*{\tauetaExp}{\tau_\eta^{\sub{exp}}}
\newcommand*{\epsN}{\varepsilon_{\sub{n}}}

\newcommand*{\vp}{\bm{v}_\sub{p}}
\newcommand*{\vpEff}{\bm{v}_\sub{eff}}
\newcommand*{\xp}{\bm{x}_\sub{p}}
\newcommand*{\Stokes}{\text{St}}

\newcommand*{\Rey}{\text{Re}}  
\newcommand*{\ReyN}{\Rey_\sub{n}}
\newcommand*{\ReyS}{\Rey_\sub{s}}

\newcommand*{\Ptwo}{P_2}
\newcommand*{\Dtwo}{\mathcal{D}_2}

\newcommand*{\Tlambda}{T_\lambda}
\newcommand*{\HeFour}{$^{4}$He}

\newcommand*{\enstrophy}{\mathcal{W}}
\newcommand*{\enstrophyP}{\enstrophy^{\text{p}}}

\newcommand*{\angstrom}{\,\text{\AA}}
\newcommand*{\kelvin}{\,\text{K}}
\newcommand*{\micron}{\,\ensuremath{\mu}\text{m}}

\begin{document}

\title{Inhomogeneous distribution of particles in co-flow and counterflow quantum turbulence}
\author{Juan Ignacio Polanco}
\author{Giorgio Krstulovic}
\affiliation{%
  Université Côte d'Azur, Observatoire de la Côte d'Azur, CNRS,
  Laboratoire Lagrange, Nice, France
}


\begin{abstract}
Particles are today the main tool to study superfluid turbulence and visualize
quantum vortices. In this work, we study the dynamics and the
spatial distribution of particles in co-flow and counterflow superfluid helium
turbulence in the framework of the two-fluid Hall-Vinen-Bekarevich-Khalatnikov (HVBK)
model. We perform three-dimensional numerical simulations of the HVBK equations
along with the particle dynamics that depends on the motion of both fluid
components.
We find that, at low temperatures, where the superfluid mass fraction dominates,
particles strongly cluster in vortex filaments regardless of their physical
properties.
At higher temperatures, as viscous drag becomes important and the two components
become tightly coupled, the clustering dynamics in the coflowing case approach
those found in classical turbulence,
while under strong counterflow, the particle distribution is dominated by the
quasi-two-dimensionalization of the flow.

\end{abstract}

\maketitle

\section{Introduction}

Turbulence has fascinated physicists and mathematicians for centuries, and is one
of the oldest yet still unsolved problems in physics. In a turbulent fluid,
energy injected at large scales is transferred
towards small scales in a cascade process~\cite{Frisch1995}. At small
scales, a turbulent fluid develops strong velocity gradients resulting in the
appearance of vortex filaments~\cite{shePumirSiggia1990intermittent}. Such
vortices have an important counterpart in turbulent quantum fluids, such as
superfluid helium and Bose-Einstein condensates (BECs) made of dilute alkali
gases. 

At finite temperatures, a quantum fluid consists of two immiscible components: a
superfluid with no viscosity, and
a normal fluid described by the Navier-Stokes equations. In the case where the
mean relative velocity of these two components is non-zero, the two-fluid description
leads to a turbulent state with no classical analogous known as counterflow
turbulence~\cite{donnelly1991quantized}.
Such out-of-equilibrium state is typically produced by imposing a
temperature gradient in a channel~\cite{VinenCounterflowExp,barenghi2014introduction}.
Another defining property of superfluids is that the circulation around
vortices is quantized.
Such objects, known as quantum vortices, have been the subject of extensive
experimental studies since the early discovery of superfluidity. Rectilinear quantum vortices were first photographed at the intersection with helium-free surfaces in 1979~\cite{YarmchukVortPhoto}. There is a renewed interest since 2006, when they were
first visualized in superfluid helium using hydrogen particles~\cite{Bewley2006}.
Further progress on particle tracking methods has enabled the
observation of quantum vortex reconnections~\cite{BewleyReconnectionPNAS}
and Kelvin waves~\cite{Fonda25032014},
as well as unveiling the differences between classical and quantum
turbulence~\cite{PaolettiVelStatQuantum,LaMantia2014a}.

Particles have been also actively used to study vortex dynamics in classical
fluids~\cite{Toschi2009}.
Particle inertia generally leads to a non-uniform spatial distribution of particles
in turbulent flows~\cite{Maxey1987}.
Light particles such as bubbles in water become trapped in
vortices allowing their visualization~\cite{DoudyCouderBrachet91Bubbles},
while heavy particles tend to escape from them~\cite{Eaton1994}.
In quantum turbulence, the situation is more complex since particles interact
with both components of the superfluid~\cite{Poole2005}.
At low temperatures where the normal fluid fraction is negligible, the particle
dynamics is dominated by pressure gradients leading to their trapping by quantum
vortices~\cite{donnelly1991quantized,CloseParticleNewcastle,giuriato2019interaction}.
As temperature increases, particles additionally
experience a viscous Stokes drag from the normal component.

There exist different models to describe superfluid turbulence.
At very low temperatures, the Gross-Pitaevskii equation describes well a
weakly interacting BEC, and is expected to provide a qualitative description of
superfluid helium.
In this model, vortices are by construction topological defects and their
circulation is therefore quantized.
The Gross-Pitaevskii equation has been generalized to include the dynamics of
classical particles~\cite{ActiveWiniecki,ShucklaSticking}, and has been used to
study particle trapping by quantum vortices~\cite{giuriato2019interaction} and
the particle-vortex interaction once particles are
trapped~\cite{giuriato2019trapped}.
A second approach is the vortex filament method, where each vortex line advects
each other through Biot-Savart integrals~\cite{SchwartzVFM}.
This method has also been adapted to describe the
interaction of particles and vortices~\cite{CloseParticleNewcastle,ParticlesNewcastle}.
Finally, a third kind of model is given
by the coarse-grained Hall-Vinen-Bekarevich-Khalatnikov (HVBK) equations~\cite{donnelly1991quantized}.
This approach is well adapted to describe the large-scale motion of a turbulent
superfluid at finite temperature, although the quantum nature of vortices is
lost.
In particular, it has been recently used to study co-flow and counterflow
turbulence~\cite{Biferale2018,Biferale2019a}.

Liquid helium experiments commonly use solid hydrogen or
deuterium particles with typical diameters of a few
microns~\cite{Svancara2019,Mastracci2019}. Although such particles are much
larger than the vortex core size $a_0 \approx 1\angstrom$, it is
expected that they do not disturb much the large scales of the superfluid.
In this work, we investigate the dynamics of particles in three-dimensional
co-flow and counterflow quantum turbulence by performing
direct numerical simulations of the HVBK model. In particular, we study how well
particles sample the different regions of the flow and how they cluster on
vortices depending on their physical properties.

\section{Governing equations}

\subsection{Coarse-grained HVBK model}\label{sub:HVBK_model}

\begin{figure}[tb]
  \centering
  \includegraphics[]{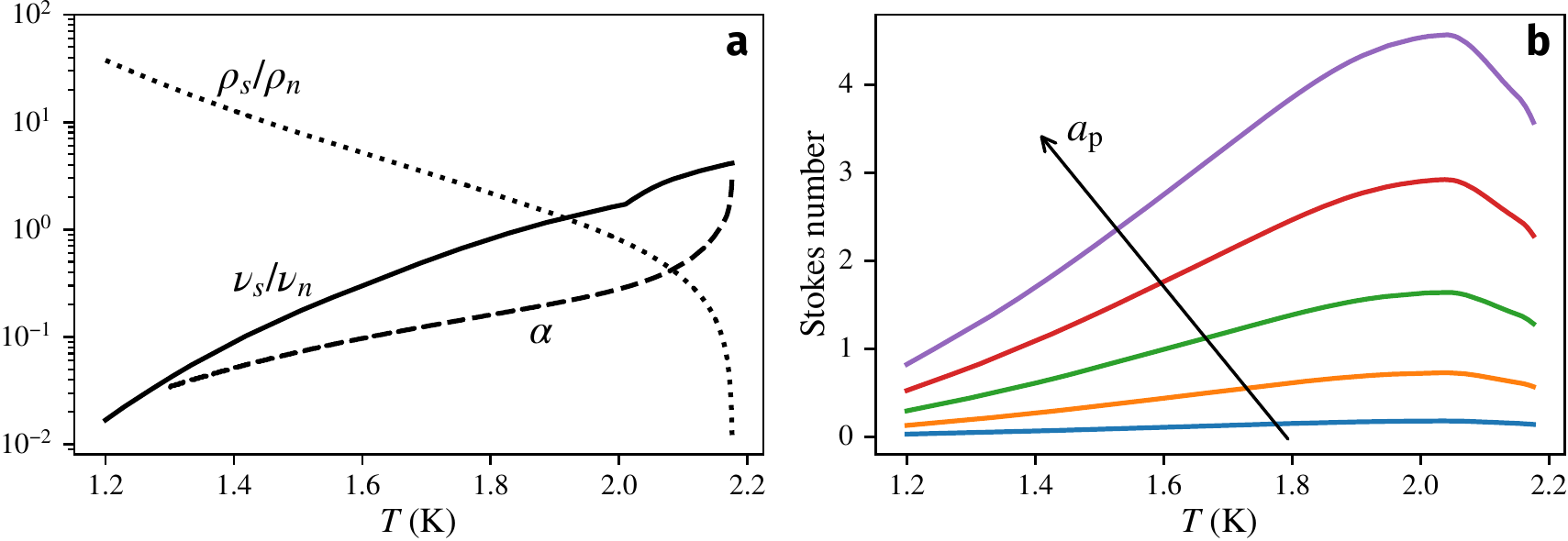}
  \caption{%
    (a) Temperature-dependent properties of superfluid \HeFour{}.
    Mutual friction coefficient $\alpha$
    from~\cite{Donnelly1998},
    density ratio $\rhos / \rhon$ and
    viscosity ratio $\nus / \nun$.
    (b) Dependence of Stokes numbers $\Stokes = \taup / \tauetaN$ on temperature.
    Stokes numbers are estimated for spherical solid hydrogen particles
    ($\rhop / \rho \approx 0.6$~\cite{LaMantia2014a}) with
    diameters ranging from 2 to 10\micron{},
    using $\tauetaN = {(\rho / \rhon)}^{1/2} \, \tauetaExp$ and
    $\tauetaExp = 0.1$\,ms~\cite{Rousset2014}.
  }\label{fig:helium_properties}
\end{figure}

We consider the dynamics of turbulent superfluid helium at finite temperature
driven by the HVBK equations,
describing the flow at scales larger than the mean distance between vortices.
At these scales, the quantum vortex dynamics can be approximated by a
coarse-grained superfluid velocity field $\vs$, which interacts with the viscous
normal component $\vn$ via two coupled Navier-Stokes equations,
\begin{align}
  \label{eq:HVBK_first}
  \dertt{\vn} +
  \vn \cdot \gradient \vn
  &=
  -\frac{1}{\rhon} \gradient \pn +
  \nun \laplacian \vn -
  \frac{\rhos}{\rhon} \mutualFriction +
  \forcing{n},
  \\
  \label{eq:HVBK_vs}
  \dertt{\vs} +
  \vs \cdot \gradient \vs
  &=
  -\frac{1}{\rhos} \gradient \ps +
  \nus \laplacian \vs +
  \mutualFriction +
  \forcing{s},
  \\
  \label{eq:HVBK_last}
  \gradient \cdot \vn
  =
  \gradient & \cdot \vs = 0,
  \quad
  \mutualFriction = \alpha \Omega_0 (\vn - \vs).
\end{align}
The total density of the fluid is $\rho = \rhon + \rhos$. The normal fluid
viscosity $\nun$ is related to the helium dynamic viscosity $\mu$ by
$\nun=\mu/\rhon$.
The two fluids are coupled through the mutual friction force
  $\mutualFriction$ that originates from the scattering of the excitations
  constituting the normal fluid component on quantum vortices. To be included in
  the HVBK dynamics, this microscopic process has to be averaged on the relevant
  scales (for a detailed discussion see Ref.~\cite{Biferale2019a}). A number of
  models have been proposed to estimate this characteristic time scale for the
  HVBK description. In general, it is proportional to the temperature dependent mutual friction coefficient $\alpha$
(see Fig.~\ref{fig:helium_properties}a) and to a characteristic superfluid
vorticity $\Omega_0$. The frequency $\Omega_0$ is in principle proportional to
the vortex line density and to the quantum of circulation.
As in Ref.~\cite{Biferale2018},
we estimate it as $\Omega_0^2 = \me{|\ws|^2} / 2$,
where $\ws = \gradient \times \vs$ is the superfluid vorticity,
and $\me{\cdot}$ denotes a space average. When there is a very strong counterflow, this superfluid vorticity-based estimate
may underestimate the mutual friction frequency.
In this case, one can instead take $\Omega_0$ as an external control parameter
depending on the particular flow~\cite{Biferale2019d}.

The two velocity fields are stirred by independent large-scale Gaussian random
forces $\forcing{s}(\xvec)$ and $\forcing{n}(\xvec)$ of unit variance.
In the present formulation, a mean counterflow velocity
$\UcfVec = \me{\vn} - \me{\vs}$ may be optionally imposed by
setting the average forces to
$\me{\forcing{s}} = -\alpha \Omega_0 \UcfVec$ and
$\me{\forcing{n}} = (\rhos / \rhon) \alpha \Omega_0 \UcfVec$.
In Eq.~\eqref{eq:HVBK_vs}, the effective superfluid viscosity $\nus$ models the
small-scale physics not resolved by the HVBK equations, including energy
dissipation due to quantum vortex reconnections and Kelvin wave excitation.
The values of the effective viscosity are taken from the model described in
Refs.~\cite{Vinen2002,Boue2015}.
The viscosity ratio $\nus / \nun$ resulting from this model is shown in
Fig.~\ref{fig:helium_properties}(a).

\subsection{Inertial particles in the HVBK model}

Particles in superfluid helium experience a Stokes drag associated to the
viscosity of the normal fluid, while
also feeling the pressure gradient force from both fluid components.
Particles are considered to be much smaller than the Kolmogorov scales of the
flow.
Hence finite-size effects can be neglected,
as well as the action of particles on the flow,
since any disturbance of the flow is immediately damped.
The Basset history term is also neglected.
The equations governing the particle dynamics then read~\cite{Poole2005,Sergeev2009}
\begin{gather}
  \label{eq:particle_equation}
  \dert{\vp}
  =
  \frac{1}{\taup}
  \left(\vn(\xp) - \vp\right) +
  \beta
  \left(
    \frac{\rhon}{\rho} \An +
    \frac{\rhos}{\rho} \As
  \right)
  \\
  \label{eq:taup_and_beta}
  \taup =
  \frac{\ap^2}{3 \beta \nu},
  \qquad
  \beta = \frac{3\rho}{2\rhop+\rho},
\end{gather}
where $\rhop$ is the particle density and $\ap$ its radius,
and $\DD / \DD t$ are the corresponding material derivatives.
The density parameter $\beta$ accounts for added mass effects,
while the Stokes time $\taup$ represents the particle response time to
normal fluid fluctuations.
Note that, even though there is a viscous term in Eq.~\eqref{eq:HVBK_vs},
there is no Stokes drag resulting from the superfluid component.
Particles moving at velocities close to the speed of sound could in principle
trigger vortex nucleations, which would result in an additional effective drag.
Here we neglect such small-scale effects.

The superfluid pressure gradient term in Eq.~\eqref{eq:particle_equation},
  proportional to $\DD \vs / \DD t$, is responsible for particle attraction
  towards superfluid vortices.
  Note that the present model does not explicitly account for particles that
  become trapped by quantum vortices, whose behavior is expected to be different
  from that of untrapped particles.
  For instance, in thermal counterflow experiments, trapped particles move
  towards the heat source along with the superfluid flow, while untrapped ones
  are transported away from it by the normal component~\cite{Paoletti2008a}.
  With regard to the spatial distribution of particles,
  one can expect that accounting for trapping would further increase the
  concentration of particles in superfluid vortices compared to the present
  model.


The Stokes number $\Stokes = \taup / \tauetaN$ quantifies the particle
inertia.
Here the Kolmogorov time scale associated to the normal component is
$\tauetaN = {(\nun / \epsN)}^{1/2}$,
where $\epsN$ is the mean energy dissipation rate of the normal fluid.
In the limit $\Stokes \to 0$,
particles behave as perfect tracers of the normal component.
In the opposite limit $\Stokes \to \infty$ the particle motion is ballistic and
not modified by turbulent fluctuations.
The Stokes numbers of micrometer-sized hydrogen particles based on dissipation
measurements in the SHREK experiment~\cite{Rousset2014} are estimated in
Fig.~\ref{fig:helium_properties}(b).
Remarkably, the temperature dependence of $\Stokes$ for fixed particle parameters
$(\ap, \rhop)$ is non-monotonic due to the variation of helium properties with
temperature,
and presents a maximum value at $T \approx 2.04\kelvin$.

As noted above, the present model is valid in the limit of small particle size
compared to the Kolmogorov scale of the normal fluid.
In addition, particles should be in principle smaller than the mean inter-vortex distance,
so that they do not interact strongly with quantized vortices, and do not
get often trapped by them~\cite{Sergeev2009}.
Consistently with the HVBK approach, which does not explicitly account for
quantized vortex dynamics, Eq.~\eqref{eq:particle_equation} should be
interpreted as describing the coarse-grained particle dynamics, neglecting the
physics at smaller scales.
Whether such small-scale phenomena have an impact on the coarse-grained particle
dynamics is a challenging question that can only be answered by confronting this
model to new experimental results.
Note that, in recent superfluid \HeFour{} experiments, the inter-vortex distance is of order
$10\micron$~\cite{barenghi2014introduction,Roche2007,Rousset2014},
comparable both to the Kolmogorov scales and to the typical size of hydrogen
particles in experiments.

\subsection{Numerical procedure}

We investigate the spatial distribution of inertial particles in superfluid
\HeFour{}
by numerically solving the HVBK
equations~(\ref{eq:HVBK_first}--\ref{eq:HVBK_last}) in a triply periodic box
using a parallel pseudo-spectral
code (see~\cite{Homann2009} for details).
Point particles are randomly initialized in the domain,
and their trajectories are evolved in time
until the system reaches a statistically steady state.
The time advancement of both particles and fields is performed using a third-order
Runge-Kutta scheme.
Fluid fields are interpolated at particle positions using fourth-order
B-splines~\cite{vanHinsberg2012}.


Simulations are performed at temperatures $T = $ 1.3, 1.9 and 2.1\kelvin.
Navier-Stokes simulations are also performed for comparison with the classical
turbulence case.
The number of collocation points in each direction is either $N =$ 256 or 512.
Both resolutions only differ on the numerical value of the viscosities $\nun$
and $\nus$ and on the resulting Reynolds numbers, but the $\nus / \nun$
ratio is kept the same.
The Reynolds numbers associated to the normal and superfluid
  components are defined as $\Rey_\alpha = \vRMS^{(\alpha)} / (\nu_\alpha k_0)$,
  where $\alpha = \{\sub{n}, \sub{s}\}$, $\vRMS^{(\alpha)}$ is the
  root-mean-square of the velocity fluctuations, and $1/k_0 = 1$ is the scale of
  the external forcing. Reynolds numbers are fixed by the resolution, as the smallest scales of the most turbulent component have to be well resolved.   
For each run, $\numParticles$ particles of a given class are tracked,
each class being defined by a set of parameters $(\ap, \rhop)$.
Simulation parameters are summarized in Table~\ref{tab:DNS_parameters}.

Two counterflow simulations (runs IV and V in Table~\ref{tab:DNS_parameters})
are performed at the temperature $T = 1.9\kelvin$ at which the two fluid components
have comparable properties.
The two runs differ on the effective mutual friction force: while the first run
uses the same estimate $\Omega_0^2 = \me{|\ws|^2} / 2$ as in the coflow runs,
the second one takes $\Omega_0$ as an external control parameter with a value $4$ times larger than steady value of the first run. The values of $\Omega_0$, normalized by $k_0 \vnRMS$ are also displayed in Table \ref{tab:DNS_parameters}. Note that effectively, the coupling between the two fluid components is stronger for run V than run IV.
As discussed in Sec.~\ref{sub:HVBK_model}, this is to account for a likely
underestimation of the mutual friction intensity by the superfluid
vorticity-based estimate.
This also allows to clarify the effect of the mutual friction on particle
concentration statistics.
In both cases, the imposed mean counterflow velocity is
$\Ucf / \vnRMS \approx 5$.

\begin{table}[htb]
  \caption{%
    Simulation parameters.
    NS denotes Navier-Stokes simulations.
    (See text for definitions.)
  }\label{tab:DNS_parameters}
  \newcommand*{\ct}[1]{\multicolumn{1}{c}{#1}}  
  \begin{ruledtabular}  
  \begin{tabular}{lccclrlllrrc}
    Run & $T$ (K) & $\Ucf / \vnRMS$ & $N$ & \ct{$\alpha$} &
    \ct{$\Omega_0 / (k_0 \vnRMS)$} & \ct{$\rho_s / \rho$} &
    \ct{$\rho_n / \rho$} & \ct{$\nu_s / \nu_n$} & $\ReyN$ & $\ReyS$ & $\numParticles / 10^6$ \\
    \midrule
    I    & 1.3 & 0.0 & 256 & 0.034 & 8.7  & 0.952 & 0.048 & 0.043 & 28   & 707  & 2.0 \\
    II   &     & 0.0 & 512 &       & 14.2 &       &       &       & 59   & 1479 & 3.2 \\
    III  & 1.9 & 0.0 & 256 & 0.206 & 7.9  & 0.574 & 0.426 & 1.25  & 632  & 516  & 2.0 \\
    IV   &     & 4.3 & 256 &       & 2.3  &       &       &       & 592  & 426  & 0.4 \\
    V    &     & 5.6 & 256 &       & 11.2 &       &       &       & 447  & 386  & 0.4 \\
    VI   &     & 0.0 & 512 &       & 11.2 &       &       &       & 1299 & 1053 & 3.2 \\
    VII  & 2.1 & 0.0 & 256 & 0.481 & 7.4  & 0.259 & 0.741 & 2.5   & 695  & 268  & 0.4 \\
    VIII &     & 0.0 & 512 &       & 9.8  &       &       &       & 1332 & 515  & 3.2 \\
    IX   & NS  & 0.0 & 256 & 0.0   & --   & 0.0   & 1.0   & --    & 780  & --   & 0.4 \\
    X    &     & 0.0 & 512 &       & --   &       &       &       & 1639 & --   & 3.2 \\
  \end{tabular}
  \end{ruledtabular}
\end{table}

\begin{figure}[tb]
  \centering
  \includegraphics[width=\columnwidth]{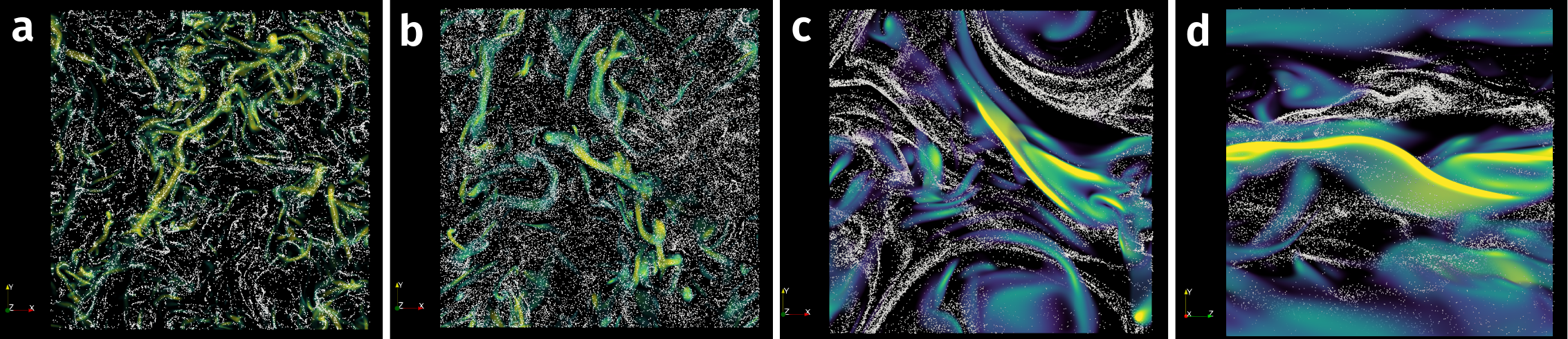}
  \caption{%
    Quasi-two-dimensional slices of the instantaneous particle distribution
    for $\Stokes = 1$ and $\rhop / \rho = 0.7$ ($\beta = 1.25$).
    (a) $T = 1.3\kelvin$ coflow, run I;
    (b) $T = 1.9\kelvin$ coflow, run III;
    (c--d) $T = 1.9\kelvin$ counterflow, run IV.
    In (c), the counterflow direction is normal to the figure.
    In (d), the counterflow is directed along the horizontal axis.
    Colors represent regions of high superfluid vorticity.
  }\label{fig:particles_visu}
\end{figure}

\section{Spatial distribution of particles}

To illustrate the effect of temperature on particle clustering,
we show in Fig.~\ref{fig:particles_visu} the instantaneous particle distribution
obtained from different simulations.
Particle parameters are $\Stokes = 1$ and $\rhop / \rho = 0.7$, comparable to those
typically found in experiments (see Fig.~\ref{fig:helium_properties}b).
In turbulent coflow at $T = 1.3\kelvin$ (Fig.~\ref{fig:particles_visu}a),
particles form quasi-one-dimensional clusters that are often aligned with
superfluid vortex filaments.
At higher temperatures ($T = 1.9\kelvin$, Fig.~\ref{fig:particles_visu}b),
the particle distribution is more uniform,
although regions of high particle concentration are still clearly visible.
When a mean counterflow is imposed at the same temperature
(Fig.~\ref{fig:particles_visu}(c) and (d)), the particle distribution is dominated by the
formation of large-scale vortices elongated along the
counterflow direction.
For the chosen set of parameters, particles tend to escape from such vortices
and concentrate in large-scale structures.


Particles in classical turbulence are known to form fractal
clusters at distances smaller than the dissipative scale of the
flow~\cite{Gustavsson2016}.
A measure of fractal clustering is the correlation dimension
$\Dtwo$~\cite{Grassberger1983,Bec2007b},
estimated as the small-scale power law scaling of the
probability $\Ptwo(r)$ of finding two particles at a distance smaller than $r$
(i.e.\ $\Ptwo(r) \sim r^{\Dtwo}$ for $r$ small).
In three dimensions, $\Dtwo = 3$ indicates that particles are uniformly
distributed in space, while smaller values are evidence of fractal clustering.

\begin{figure}[t]
  \centering
  \includegraphics{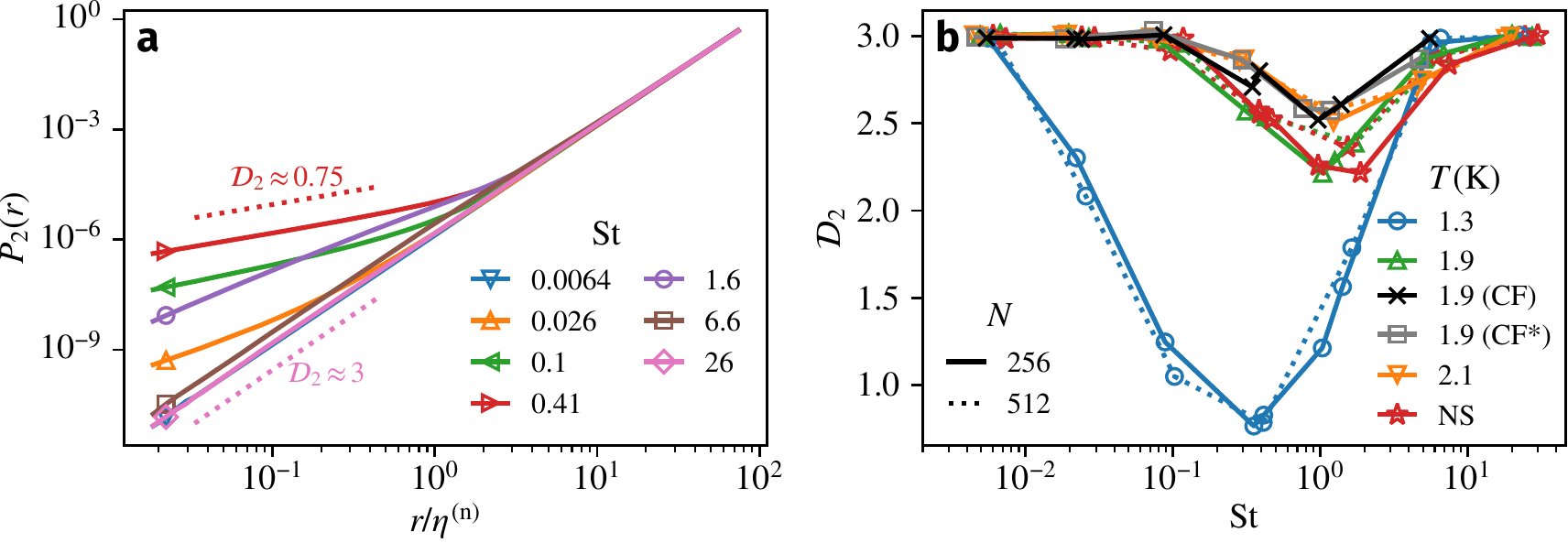}
  \caption{%
    Particle concentration at constant density $\rhop / \rho = 0.7$
    ($\beta = 1.25$).
    (a) Separation probability $\Ptwo(r)$ for $T = 1.3\kelvin$ (run II)
    and for different Stokes numbers.
    Distances are normalized by the normal component Kolmogorov length scale
    $\etaN = (\nun^3 / \epsN)^{1/4}$.
    (b) Correlation dimension $\Dtwo$ as a function of Stokes number
    for all runs.
    Different markers represent different cases.
    CF: counterflow turbulence (run IV).
    CF*: counterflow turbulence with strong mutual friction (run V).
    NS: classical turbulence (runs IX, X).
  }\label{fig:concentration_D2}
\end{figure}

We first consider particles of relative density $\rhop / \rho = 0.7$
at varying particle radius $\ap$.
The separation probability $\Ptwo(r)$ for different Stokes
numbers is shown in Fig.~\ref{fig:concentration_D2}(a) for the 1.3\kelvin{}
cases.
At small scales, the curves present a clear power law scaling,
with an exponent $\Dtwo$ that varies significantly with $\Stokes$.
At this temperature, particle clustering is maximal for $\Stokes \approx 0.4$,
which would roughly correspond to 6\micron-diameter hydrogen particles in
SHREK (Fig.~\ref{fig:helium_properties}b) or in the Prague oscillating
grid experiments~\cite{Svancara2017a}.

We plot in Fig.~\ref{fig:concentration_D2}(b) the correlation dimension $\Dtwo$
from all runs.
As in classical turbulence~\cite{Eaton1994},
for all temperatures particle clustering is maximal at Stokes numbers of order
unity.
At temperatures close to $\Tlambda$, the minimum value of $\Dtwo$ is close to
2.3, comparable to the case of heavy particles in turbulence~\cite{Bec2007a}.
In particular, both counterflow cases at $T = 1.9\kelvin$ display a maximum
clustering at
$\Stokes \approx 1$, similarly to the coflow runs at the same temperature.
The two counterflow curves nearly collapse, suggesting that there is no
significant effect of the mutual friction intensity on $\Dtwo$.
As anticipated from Fig.~\ref{fig:particles_visu}, particle
clustering changes dramatically in turbulent coflow at lower temperatures.
At $T = 1.3\kelvin$, the minimum value of $\Dtwo$ decreases to 0.75,
indicating that particles become concentrated in worm-like structures such as
those seen in Fig.~\ref{fig:particles_visu}(a).

To understand the above observations, we consider the particle equation of
motion in
the small Stokes number limit ($\taup \ll \tauetaN$).
In this case, particles follow an effective
compressible velocity field
$\vpEff(\xvec, t)$~\cite{Maxey1987,Balkovsky2001}.
From Eq.~\eqref{eq:particle_equation},
this field writes
$\vpEff \approx
\vn +
\taup \left( \beta \frac{\rhon}{\rho} - 1 \right) \An +
\taup \, \beta \frac{\rhos}{\rho} \As$.
Taking its divergence, one finds
\begin{equation}
  \label{eq:divergence_vp}
  \frac{1}{\taup} \gradient \cdot \vpEff \approx
  \left( \beta \frac{\rhon}{\rho} - 1 \right) (\Sn^2 - \On^2) +
  \beta \frac{\rhos}{\rho} (\Ss^2 - \Os^2),
\end{equation}
where $\Ss$, $\Sn$, $\Os$ and $\On$ are the norms of the strain-rate and
rotation-rate tensors of the two fluids.

In the classical limit where $\rhos = 0$,
Eq.~\eqref{eq:divergence_vp} indicates that light particles ($\beta > 1$) tend
to concentrate in vorticity-dominated regions (where $\On > \Sn$),
while heavy particles ($\beta < 1$) accumulate in strain-dominated
regions~\cite{Balachandar2010}.
For neutral particles ($\beta = 1$), the effective velocity field is
incompressible and no preferential concentration is expected.

The classical picture changes in low-temperature \HeFour{} when $\rhos \gg \rhon$.
In this case, Eq.~\eqref{eq:divergence_vp} becomes
${\taup}^{-1} \gradient \cdot \vpEff \approx
- (\Sn^2 - \On^2) + \beta (\Ss^2 - \Os^2)$,
implying that the remaining normal component acts on the particle dynamics only
through the Stokes drag.
Due to its higher viscosity, the normal velocity field is smoother (has
weaker gradients), hence in general
$|\Ss^2 - \Os^2| \gg |\Sn^2 - \On^2|$.
As a consequence, for $\beta$ of order unity,
the superfluid term dominates, and thus
particles cluster in regions of high superfluid vorticity.
We stress that this behavior is unique to quantum turbulence,
since the absence of superfluid drag on the particles implies
that there is no force counteracting the dominant effect of the superfluid
pressure gradient.

In the opposite limit $T \to \Tlambda$,
the superfluid fraction vanishes and the classical behavior discussed above is
recovered.
More interesting is the intermediate case where the two fluid densities
and viscosities are similar, at $T \approx 1.9\kelvin$.
In this case, in the absence of a mean counterflow, the two velocity
fields are tightly coupled even at the smallest flow
scales~\cite{Biferale2018}.
Hence, $\Sn \approx \Ss$ and $\On \approx \Os$, and the clustering behavior
predicted by Eq.~\eqref{eq:divergence_vp} falls back to the classical case.
This does not apply in the counterflow case,
where the normal and superfluid motions are decorrelated at the small scales
along the counterflow direction~\cite{Biferale2019a}.

\begin{figure}[t]
  \centering
  \includegraphics{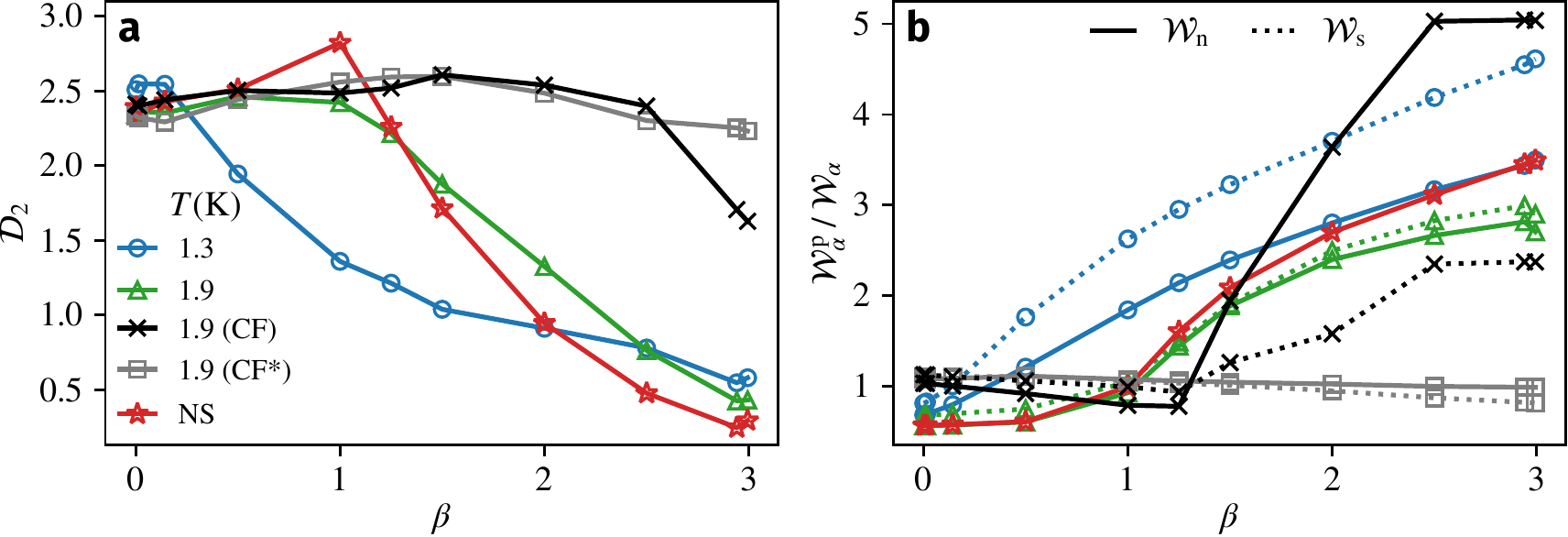}
  \caption{%
    Clustering as a function of particle density for $\Stokes = 1$.
    (a) Correlation dimension $\Dtwo$.
    (b) Relative enstrophy sampled by the particles,
    $\enstrophyP_\alpha / \enstrophy_\alpha$,
    for $\alpha = \sub{n}, \sub{s}$.
    Solid lines, normal fluid enstrophy;
    dotted lines, superfluid enstrophy.
  }\label{fig:concentration_vs_beta}
\end{figure}

To support the above predictions and to extend our results to different
particle densities, Fig.~\ref{fig:concentration_vs_beta}(a) shows $\Dtwo$
as a function of the density parameter $\beta$, for $\Stokes = 1$.
In the classical case, heavy particles concentrate in planar
structures ($\Dtwo \gtrsim 2$) while light particles form localized linear
clusters ($\Dtwo \lesssim 1$), consistently with previous
findings~\cite{Calzavarini2008a}.
For neutral particles, $\Dtwo$ strongly decreases at the lowest temperature,
suggesting the formation of linear clusters.
The above discussion suggests that these clusters form in high superfluid
vorticity regions.
This is verified in Fig.~\ref{fig:concentration_vs_beta}(b), where the relative
enstrophy sampled by the particles, $\enstrophyP_\alpha / \enstrophy_\alpha$
(with $\alpha = \sub{n}, \sub{s}$), is plotted.
Here, $\enstrophy_\alpha = \me{|\vort_\alpha|^2}$ is the enstrophy of a given
fluid component (Eulerian average), and $\enstrophyP_\alpha = \me{|\vort_\alpha(\xp)|^2}$ where
the average is over particle positions.
For $T = 1.3\kelvin$, neutral and light particles preferentially sample high
superfluid vorticity regions.

We finally discuss the counterflow case.
Contrarily to classical and coflowing 3D turbulence, here the two-fluid motion
is characterized by
large-scale vortices elongated along the counterflow direction and
quasi-two-dimensionalization of the flow (as seen in
Fig.~\ref{fig:particles_visu}(c-d)),
while the small scales are strongly damped due to mutual
friction~\cite{Biferale2019a}.
In contrast with the coflowing case, where the results are strongly dependent
on the density parameter $\beta$, here particles form clusters of dimension
$\Dtwo \approx 2$ almost regardless of their density
(Fig.~\ref{fig:concentration_vs_beta}a).
This is explained by particles concentrating in sheet-like structures as a
result of the two-dimensionalization of the flow.
The observed particle organization may explain why quantum vortices are not
clearly visualized by particles in some counterflow
experiments~\cite{LaMantia2014}.
Except for the case of very light particles, the mutual friction intensity has
virtually no influence on $\Dtwo$, consistently with the observations from
Fig.~\ref{fig:concentration_D2}(b).
This is however not the case for the relative enstrophy sampled by the particles
(Fig.~\ref{fig:concentration_vs_beta}b), which displays a striking variation
with the mutual friction frequency $\Omega_0$.
In the low $\Omega_0$ case, light particles tend to cluster in regions of very
high normal fluid vorticity,
while this is not the case when $\Omega_0$ is increased.
This is a consequence of the change of Eulerian fields with the mutual
friction intensity.
A strong mutual friction results in weaker enstrophy fluctuations in the flow (data not shown).
Furthermore, mutual friction suppresses the velocity fluctuations in the counterflow
direction~\cite{Biferale2019d}, enhancing the two-dimensionalization of the
flow and thus the formation of vortex sheets that drive particle clustering.
This finally explains why the correlation dimension $\Dtwo$ remains close to two
when the mutual friction frequency is increased.

\section{Summary}

We have studied the spatial organization of inertial particles in
the HVBK framework for superfluid helium.
In the absence of a mean counterflow,
the most striking difference with classical fluids is observed at low
temperatures when the superfluid mass fraction is dominant.
In this case, particles are attracted towards high superfluid vorticity regions
regardless of their density relative to the fluid, thus forming
quasi-one-dimensional clusters.
This attraction is explained by the dominant effect of the superfluid
pressure gradient on the particles.
At higher temperatures, as the two fluid components become strongly coupled by
mutual friction, the classical turbulence behavior is recovered.
Namely, light particles concentrate in vortex filaments, while heavy particles
are expelled from them.
Finally, in the presence of a strong counterflow, the clustering dynamics is
governed by the two-dimensionalization of the velocity fields and the formation
of large-scale vortex columns or sheets, which either attract or repel particles as a
function of the particle density and/or inertia.
In this case, particles cluster in quasi-2D structures almost
regardless of their density and of the imposed mutual friction intensity.

\begin{acknowledgments}
  This work was supported by the Agence Nationale de la Recherche through the
  project GIANTE ANR-18-CE30-0020-01.
  Computations were carried out on the Mésocentre SIGAMM hosted at the
  Observatoire de la Côte d'Azur, and on the Occigen cluster hosted at CINES
  through the GENCI allocation A0072A11003.
\end{acknowledgments}

\bibliography{Concentration}

\end{document}